\documentclass[runningheads]{llncs}
\AtBeginDocument{%
  \providecommand\BibTeX{{%
    \normalfont B\kern-0.5em{\scshape i\kern-0.25em b}\kern-0.8em\TeX}}}

\usepackage{color}
\usepackage{algorithm}
\usepackage{algorithmic}
\usepackage{graphicx}
\usepackage{amsmath}
\usepackage{multirow}
\usepackage{balance} 
\usepackage{mathrsfs}

\begin{document}
\title{An Adaptive Framework of Geographical Group-Specific Network on O2O Recommendation}
\titlerunning{Geographical Group-Specific Network on O2O Recommendation}
%
\author{Luo Ji\inst{1}\orcidID{0000-0002-2484-5345} \and
Jiayu Mao\inst{2}\protect\footnote{The first two authors contributed equally to this research.}\orcidID{0009-0003-0890-5904} \and
Hailong Shi\inst{3}\protect\footnote{Corresponding author.} \and
Qian Li\inst{2} \and
Yunfei Chu\inst{1}\orcidID{0000-0002-3033-2984} \and
Hongxia Yang\inst{1}\orcidID{0000-0002-0580-9728}
}


\authorrunning{L. Ji et al.}
%
\institute{DAMO Academy, Alibaba Group,
\\ \email{\{jiluo.lj,fay.cyf,yang.yhx\}@alibaba-inc.com}
\and
Alibaba Group,
\email{\{jiayumao.mjy,lq167324\}@alibaba-inc.com}\\
\and
Institute of Microelectronics, Chinese Academy of Sciences,
\\ \email{shihailong2010@gmail.com} 
}
\maketitle              
\begin{abstract}
Online to offline recommendation strongly correlates with the user and service's spatiotemporal information, therefore calling for a higher degree of model personalization. The traditional methodology is based on a uniform model structure trained by collected centralized data, which is unlikely to capture all user patterns over different geographical areas or time periods. To tackle this challenge, we propose a geographical group-specific modeling method called GeoGrouse, which simultaneously studies the common knowledge as well as group-specific knowledge of user preferences. An automatic grouping paradigm is employed and verified based on users' geographical grouping indicators. Offline and online experiments are conducted to verify the effectiveness of our approach, and substantial business improvement is achieved.

\end{abstract}

\section{Introduction}

\begin{figure}[t]
  \centering
  \includegraphics[width=0.9\linewidth]{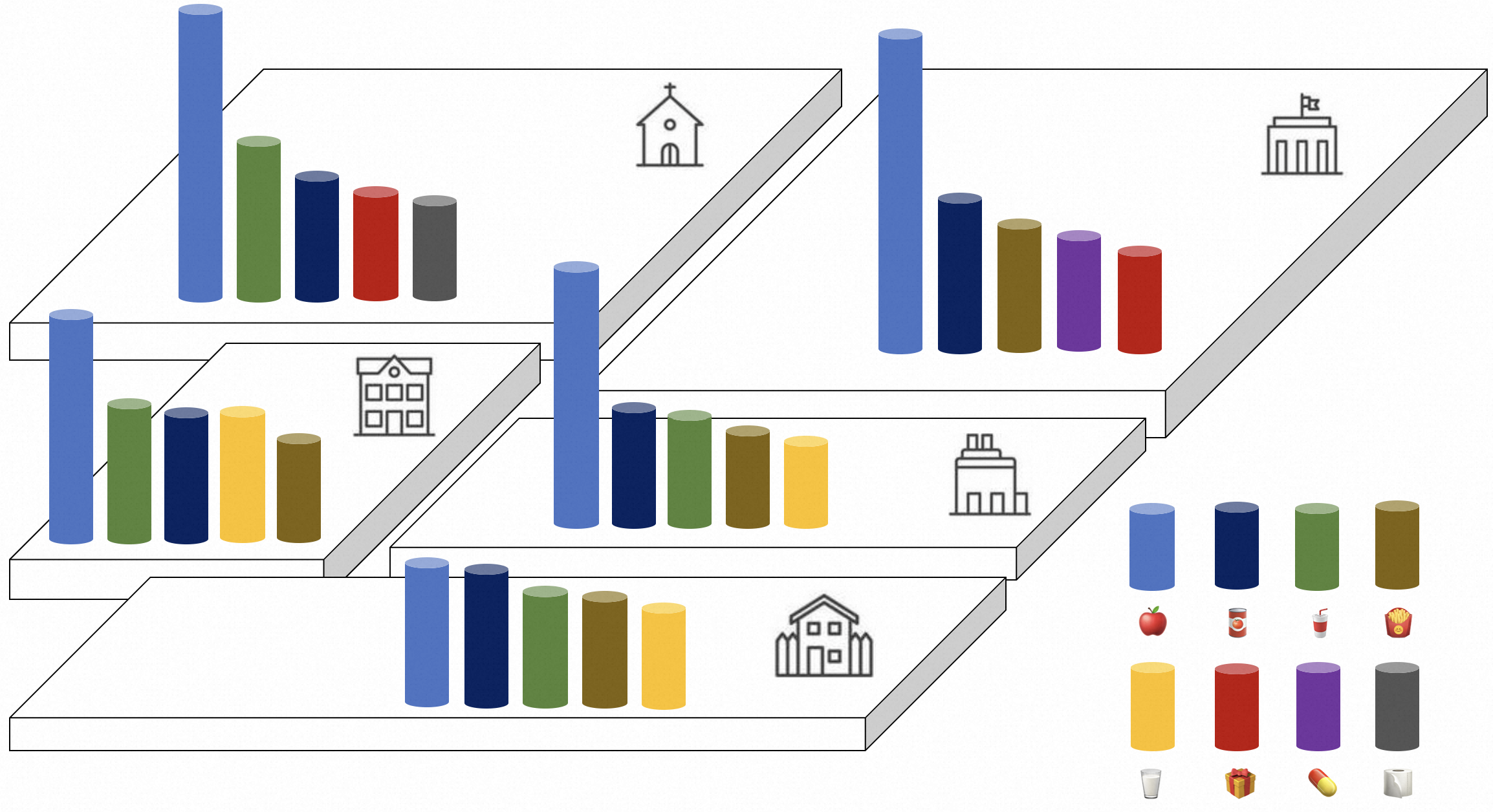}
  \caption{Geographic influence on order distributions of an O2O retail delivery platform. Top-Five item categories and their fractions are exhibited for five different functional regions (residence, working, business, education, and hospital).}
  \label{fig:example}
  \vspace{-3.5mm}
\end{figure}

Online to offline (O2O) platforms such as Uber and Meituan map online users with offline service providers on users' smartphones. This mapping is naturally geographically and temporal influenced, which is significantly different from traditional e-commerce platforms like Amazon/Taobao. Examples of this spatiotemporal influence include 1) for a specific user, only services within his/her adjacent area are applicable candidates according to the order fulfillment possibility, resulting in an extremely sparse user-item interaction matrix (sparsity inevitably happens when user and item from different areas); 2) users' interests may vary dramatically in different time periods (\textit{e.g.} food orders in the morning or evening; traveling options in workdays or weekends) ; 3) users from different geographical areas could have varied food tastes and therefore distinct behavior patterns (see Figure \ref{fig:example} as an illustrative example). These characteristics introduce more challenges for reasonable servicing personalization with respect to user spatiotemporal information. For the conventional unified model architecture \cite{Wang2019SRS},  user data across all time periods and geographical areas are leveraged together to study a uniform model representation, which may suffer performance degradation given non-uniform data distribution as shown in Figure \ref{fig:example}. On the contrary, one can choose to train a distinct model on each different geographical area and time period, to better capture local data distributions. Nevertheless, one needs to arbitrarily determine the model granularity, and fail to capture the user behavior commonality \cite{Sheng2021STAR}. Data of each model partition is also much more sparse than the uniform framework. 



In this work, we propose a novel \textbf{Geo}graphic \textbf{Grou}p-\textbf{s}p\textbf{e}cific (GeoGrouse) model framework to tackle the aforementioned challenges, on Ele.me\footnote{\url{https://www.ele.me/}}, a world-leading O2O food delivery application. Similar to STAR architecture \cite{Sheng2021STAR}, our model includes a shared-central network, as well as group-specific networks each of which is tailored to a specific user group. During training, the central network is trained on the entire data scope to capture user commonality; while the group-specific network is deployed on the device side and provides the group-level specializations by finetuning with its corresponding group data. The user grouping indicator is determined by a trainable latent embedding function with user geographical features as input. This methodology can be generalized to different types of user grouping specifications. The main contributions of this paper include:
\begin{itemize}
\item To the best of our knowledge, this is the first time to incorporate the idea of group-specific modeling with O2O recommendation, for better personalization of spatiotemporal influences. 
\item We design an adaptive user grouping mechanism instead of arbitrary user grouping. 
\item Performances of GeoGrouse on different business indicators are verified by realistic live experiments.
\end{itemize}

\section{Method}
\label{sec:method}




\subsection{Framework}
\label{sec:framework}

\begin{figure}[t]
  \centering
  \includegraphics[width=0.9\linewidth]{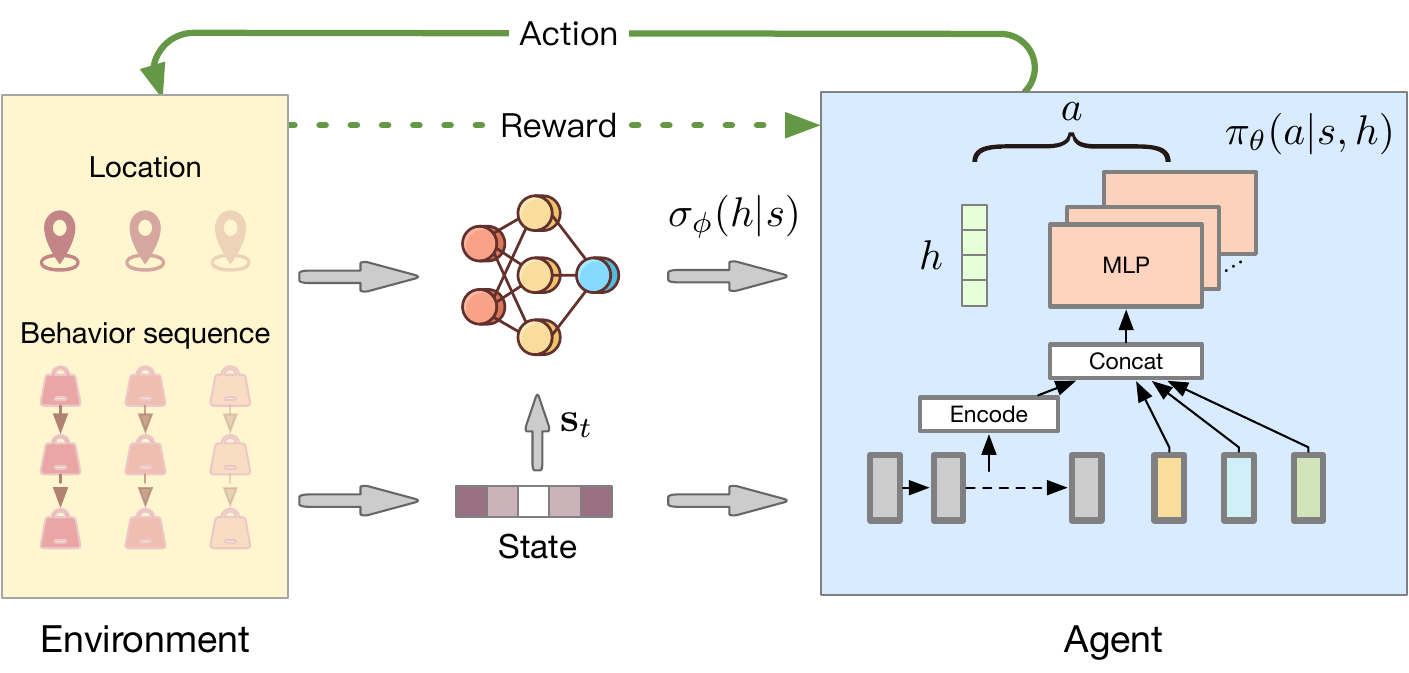}
  \caption{The framework of GeoGrouse. User states are processed with a centered module and group-specific modules, while $\sigma_{\phi}$ generates the user grouping latent variable which determines the active group-specific module.} 
  \label{fig:framework}
  \vspace{-5mm}
\end{figure}



Reinforcement Learning (RL) is an interactive learning algorithm between the agent and the environment. The agent observes the state $s$, acts with the action $a$, and receives the reward $r$ from the environment. An episode with length $t$ can be denoted as $\tau_t := \{s_0, a_0, r_0, s_1, a_1, r_1, \cdots, s_t, a_t, r_t \}$. The state transits by $\mathcal{T}(s_{t+1} \vert s_t, a_t)$.  The objective is the discounted accumulated rewards $G_t = \sum_{t}^{\infty} \gamma^t r_t$ with $\gamma \in (0, 1]$ as the discounted factor, and the agent aims to find an optimal policy $\pi(a \vert s)$ which maximizes the expected $G_t$.



Here we employ this RL framework to solve the Top-K recommendation problem, with a system configuration similar to \cite{Chen2019TopKOC}. Nonetheless, motivated by the spatial-temporal dependency of O2O, we model our policy by explicit user grouping. In this work, we further assume the distribution of states is implicitly determined by a latent grouping variable $h$, with the likelihood recognition function $\sigma(h \vert s)$. Accordingly, the original policy $\pi(a \vert s)$ becomes a latent space policy $\pi(a \vert s, h)$. Below are the detailed definitions of system variables:
\begin{itemize}
    \item \textbf{$s$}: the user profiles, historical behavior sequences, and context features including the season, weather, and geographic info (denoted by $g$).
    \item \textbf{$a$}: embedding of recommended items.
    \item \textbf{$r$}: the immediate reward obtained after a recommendation, assigned as 1 with a click or add-to-cart, and 0 otherwise.
    \item \textbf{$h$}: the grouping indicator as a learnable embedding of $g$.
\end{itemize}
%

Similar to the STAR topology \cite{Sheng2021STAR}, our policy network is a combination of one group-shared module and multiple group-specification modules. Grouping is achieved by parametric recognition model $\sigma_{\phi}(h | s)$ which is jointly learned with the parametric policy $\pi_{\theta}(a | s, h)$. We name this recommendation method as Geographic Group-Specific (GeoGrouse) network, as indicated by Figure \ref{fig:framework}. 


\subsection{Implementation of Group-Specification}
\label{sec:gs}

\begin{figure}[t]
  \centering
  \includegraphics[width=1.0\linewidth]{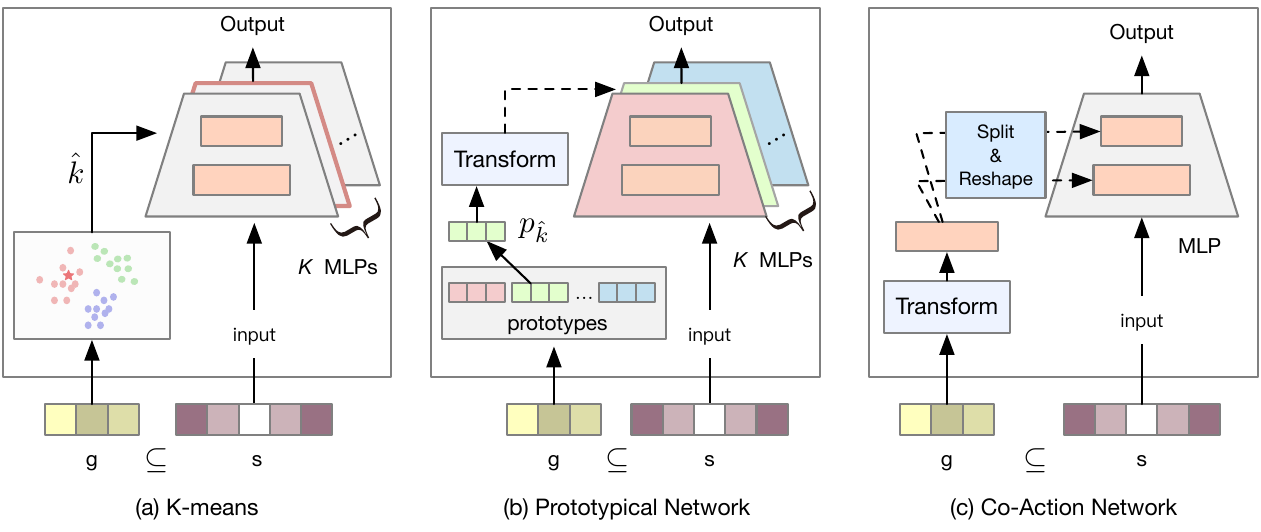}
  \caption{Comparison of Grouping Implementations. $g$ is the geography-related part among the $s$ attributes.} 
  \label{fig:group}
\end{figure}

As stated in Section \ref{sec:framework}, the policy network $\pi$ includes the group-shared module at the bottom and the group-specification module at the top. The group-shared module and the group-specification module are then denoted by 
\begin{equation*} 
    a_{s} = \text{DIN}_{\mu}(s), \quad a = \text{GS}_{\eta} (a_s, h)
\end{equation*}
in which $a_s$ is the shared part of action and DIN is the Deep Interest Network \cite{Zhou2018DIN} tower. 
The policy can then be re-expressed as $\pi_{\theta}(a \vert s, h) = \text{GS}_{\eta}(\text{DIN}_{\mu}(s), h)$ with $\theta$ as union of $\{ \mu, \eta \}$. The embedding of the grouping indicator can be further expressed as the parametric form of $h = \sigma_{\phi}(g)$. In the following subsections we propose three possible group-specification implementations of $\sigma_{\phi}(g)$ and $\text{GS}_{\eta} (a_s, h)$, with their architectural comparison shown in Figure \ref{fig:group}.


\subsubsection{K-Means.}
\label{sec:kmeans}

K-Means is a classic clustering method and is tightly correlated with MLE and EM \cite{Hu2015Thesis}. With the number of clusters $K$ as key hyper-parameter, K-Means acts as $\sigma_{\phi}$ which first learns $K$ cluster centroid $\{g_k\}_{k=1}^K$, then determine the most nearby cluster from the current $g$
\begin{equation*}
h =  \hat{k} = \arg \min_{k \in [1, K]} \lVert g - g_k \rVert_2
\end{equation*}
Then $K$ identical MLP towers are implemented to form $GS_{\eta}$ 
\begin{equation*}
    a^k = \text{MLP}_{\eta_k}(a_s), \quad k = 1, \cdots, K
\end{equation*}
then $a$ is simply the output selection of the $\hat{k}$th tower, $a = a^{\hat{k}}$ with $\eta = \{\eta_1, \cdots, \eta_k\}$. During training, $\text{MLP}_{\eta_k}$ is only trained with samples of the $k$th cluster to achieve the group-specialization.

\subsubsection{Prototypical Networks.}
\label{sec:prototype}
Similar to K-means, the prototypical method \cite{Li2020prototypical} also intrigues $K$ towers $\text{MLP}_{\eta_k}$ but in a more automatic manner. First $h$ is represented by $K$ learned prototype vectors, i.e.,  $\{{p}_k\}_{k=1}^K$, using method in \cite{Li2020prototypical}, the current optimal prototype is determined by 
\begin{equation*}
    \hat{k} = \mathop{\arg\max}\limits_{k \in [1, K]}{\text{cos}(g, {p}_k)}
\end{equation*}
where $\text{cos}(\cdot, \cdot)$ is the cosine similarity. Then for each $k$, ${p_k}$ can be further transformed to $\eta_k$ with a uniform expression
\begin{equation*}
    \eta_{k} = \text{tanh}({W}{p}_{k}+{b}),
\label{eq:t}
\end{equation*}
where ${W}$ and ${b}$ are trainable and $\eta = [{W}, {b}]$.


\subsubsection{Co-Action Network.}
\label{sec:can}

Co-Action Network (CAN) \cite{Bian2022CAN} is a feature-cross processing technique that provides an automatic manner of group specification, without the inclusion of explicit $K$ separated towers. By linearly transforming $g$ to $h$ and directly utilizing it as the weight \& bias parameter of micro-MLP tower,
\begin{align*}
    h &= L_{\phi} g, \quad a = \text{MLP}_{\eta = h} (a_s)
\end{align*}
a uniform-structured group-specification module is then obtained which can be automatically adapted to different $g$.





\subsection{Algorithm}

We approximate our solution by the famous Expectation-Maximization method (EM) \cite{Dempster1976EM}. During the Expectation stage, the latent variable is recognized by maximizing the likelihood of $\phi$ with the fixed $\theta$: 
\begin{align}
\label{eq:h_expect}
    L(\phi) &:= \log P(h \vert \tau) \sim \frac{1}{N} \sum_{s \sim \mu(\pi_{\theta})}^{|s|=N} \log q_{\phi}(h | s)
\end{align} 
On the Maximization stage, the policy parameter $\theta$ is updated given the current best estimate $\hat{h}$. Analogous to the original REINFORCE (Section 13.3 in \cite{sutton1999reinforcement}) derivation, we have
\begin{align}
\label{eq:theta_max}
    \nabla J(\theta) &\sim E_{\pi} \sum_a q_{\pi}(s, a) \nabla \pi(a \vert s, \hat{h}) = E_{\pi} [G_t \nabla \ln \pi(a \vert s, \hat{h})]
\end{align}

\section{Experiment}
\label{sec:experiment}

We launch GeoGrouse on the Ele.me platform for the retail product-instore recommendation. CAN in Section \ref{sec:can} is adopted as the default group-specification logic since it has the best experimental result. Codes have been made public\footnote{https://github.com/AaronJi/GeoGrouse}.

%

 




\subsection{Experimental Configurations}


We obtain the geographic features $g$ by concatenating embeddings of spatiotemporal features, such as city, GPS, area-of-interests (AOI), hour, and season. 

The model is trained with data extracted from 60 days' logs. The average session length is 35 while the maximum is 586.  We compare GeoGrouse with several baselines including (1) \textbf{StEN} \cite{Lin2022StEN} has state-of-the-art performance on O2O recommendation which encodes spatiotemporal information by specially designed activation and attention. (2) \textbf{DIN} (Deep Interest Network) \cite{zhou2018deep} has a local activation that captures the user interest with the target item, but with no specific spatiotemporal logic. (3) \textbf{DeepFM} \cite{guo2017deepfm} is a classical cross-feature technique for deep neural networks. 



\subsection{Offline Experiment and Sensitivity Analysis}

\begin{figure}[!h]
  \centering
  \includegraphics[width=0.8\linewidth]{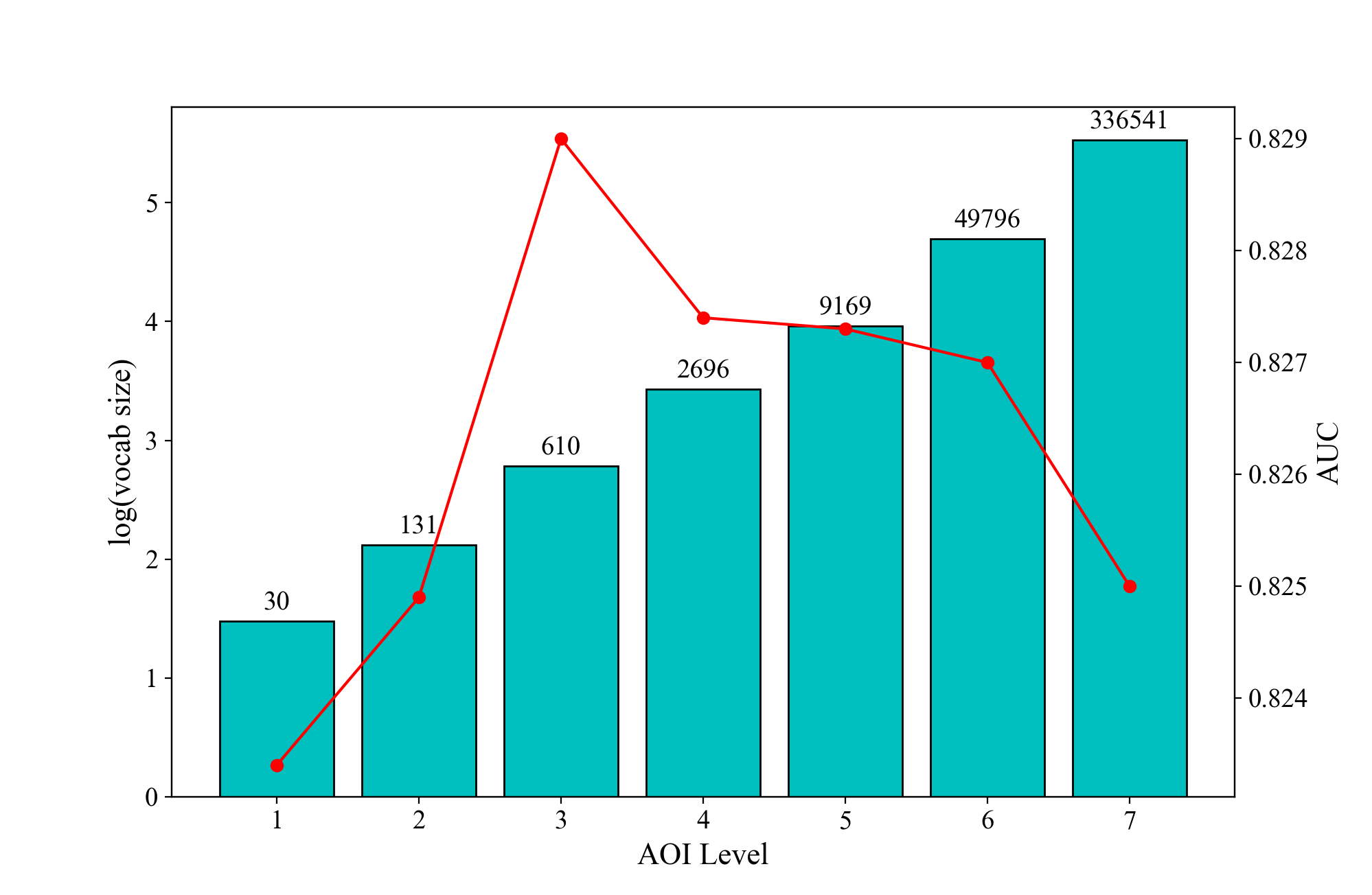}
  \caption{Sensitivity Analysis of AOI Levels.} 
  \label{fig:sensitivity}
  \vspace{-5mm}
\end{figure}

Data from the very last day is used as the test set. Experiments are repeated 10 times. Widely-used metrics such as Area Under Curve (AUC), Normalized Discounted Cumulative Gain (NDCG), and Hit Rate are used for evaluation. Table \ref{tab:offline_result} shows the offline results. GeoGrouse outperforms baselines on metrics. Among the baselines, StEN is obviously better than DIN and DeepFM, indicating the importance of spatiotemporal considerations. We also perform a sensitivity analysis of AUCs according to the choice of AOI level (and its vocabulary size), which is one of the key geographic indicators of $g$. Figure \ref{fig:sensitivity} indicates the optimal AOI level is 3 therefore we adopt this grouping granularity in formal experiments.


\begin{table}[htb]
\small
  \caption{Result of Offline Experiment}
  \vspace{-0.2in}
  \label{tab:offline_result}
  \begin{center}
  \begin{tabular}{ccccc}
    \hline
    Model & StEN & DIN & DeepFM & GeoGrouse \\ 
    \hline
    AUC & 0.820$\pm0.004$ & 0.658$\pm0.005$ & 0.778$\pm0.006$ & \textbf{0.832}$\pm0.007$ \\ 
    NDCG@3 & 0.672$\pm0.012$ & 0.504$\pm0.010$ & 0.575$\pm0.011$ & \textbf{0.674}$\pm0.012$ \\
    NDCG@5 & 0.695$\pm0.014$ & 0.536$\pm0.011$ & 0.606$\pm0.015$ & \textbf{0.696}$\pm0.015$ \\
    NDCG@10 & 0.728$\pm0.015$ & 0.583$\pm0.017$ & 0.651$\pm0.015$ & \textbf{0.730}$\pm0.017$ \\
    NDCG@20 & 0.759$\pm0.016$ & 0.627$\pm0.018$ & 0.691$\pm0.018$ & \textbf{0.760}$\pm0.017$ \\
    NDCG@50 & 0.783$\pm0.015$ & 0.665$\pm0.015$ & 0.721$\pm0.017$ & \textbf{0.784}$\pm0.018$ \\
    Hit Rate@10 & 0.959$\pm0.006$ & 0.893$\pm0.005$ & 0.932$\pm0.008$ & \textbf{0.960}$\pm0.009$ \\
  \hline
\end{tabular}
\end{center}
\vspace{-5mm}
\end{table}





\vspace{-5mm}






\subsection{Online A/B Test and Ablation Test}

The online A/B test lasts for 7 days. The key performance index (KPI) includes the click-through rate (CTR), the add-to-cart rate (ACR), the number of users with impressions (impress-UV), the number of users with clicks (click-UV), and the number of users with add-to-cart behavior (cart-UV). Due to online industrial constraints, only StEN is deployed as the live baseline. Compared with StEN and Geogrouse with group-specification of K-means and Prototypical (denoted as `GeoGrouse-K' and `GeoGrouse-P'), GeoGrouse improves all KPIs substantially as shown in Table \ref{tab:live_result}. 


\begin{table}[htb]
\small
  \caption{Result of Live Experiment. Results of GeoGrouse-K and GeoGrouse-P are relative numbers to GeoGrouse.}
  \vspace{-0.1in}
  \label{tab:live_result}
  \begin{center}
  \begin{tabular}{ccccc}
    \hline
    Model & StEN & GeoGrouse & GeoGrouse-K & GeoGrouse-P \\ %
    \hline
    CTR & 13.08\% & \textbf{13.20\%} & -0.50\%  & -0.05\% \\ 
    ACR & 9.99\% & \textbf{10.06\%} & -0.03\%  & -0.02\% \\
    impress-UV & 313,206 & \textbf{313,920} & +0.97\%  & -0.04\% \\
    click-UV & 40,980 & \textbf{41,439} & -0.81\%  & -0.43\% \\
    cart-UV & 31,277 & \textbf{31,579} & -0.67\%  & +0.03\% \\
  \hline
\end{tabular}
\end{center}
\vspace{-12mm}
\end{table}

\section{Conclusion}
\label{sec:conclusion}



In this paper, we propose a novel GeoGrouse method that applies self-adaptive user group-specification to O2O recommendation, for better personalization. Our approach is not limited to geographical factors but can be generalized to any grouping considerations. One limitation is the increased mode size due to multiple group-specific modules, which can be alleviated by split-deployment on edge devices \cite{Gong2020EdgeRec}. In the future, it would be interesting to examine the broader scope of user grouping possibilities and attempt different levels of grouping granularity.

\bibliographystyle{splncs04}
\bibliography{main}

\begin{thebibliography}{10}
\providecommand{\url}[1]{\texttt{#1}}
\providecommand{\urlprefix}{URL }
\providecommand{\doi}[1]{https://doi.org/#1}

\bibitem{Bian2022CAN}
Bian, W., Wu, K., Ren, L., Pi, Q., Zhang, Y., Xiao, C., Sheng, X.R., Zhu, Y.N.,
  Chan, Z., Mou, N., Luo, X., Xiang, S., Zhou, G., Zhu, X., Deng, H.: Can:
  Feature co-action for click-through rate prediction. In: Proceedings of the
  15th ACM International Conference on Web Search and Data Mining. WSDM '22
  (2022)

\bibitem{Chen2019TopKOC}
Chen, M., Beutel, A., Covington, P., Jain, S., Belletti, F., Chi, E.H.: Top-k
  off-policy correction for a reinforce recommender system. In: Proceedings of
  the Twelfth ACM International Conference on Web Search and Data Mining. p.
  456–464. WSDM '19 (2019)

\bibitem{Dempster1976EM}
Dempster, A., Laird, N., Rubin, D.: Maximum likelihood from incomplete data via
  the em algorithm. In: Proceedings of the Royal Statistical Society. pp. 1--38
  (1976)

\bibitem{Gong2020EdgeRec}
Gong, Y., Jiang, Z., Feng, Y., Hu, B., Zhao, K., Liu, Q., Ou, W.: Edgerec:
  Recommendation system on edge in mobile taobao. In: Proceedings of the 2020
  ACM on Conference on Information and Knowledge Management. CIKM '20 (2020)

\bibitem{guo2017deepfm}
Guo, H., Tang, R., Ye, Y., et~al.: Deepfm: a factorization-machine based neural
  network for ctr prediction. In: Proceedings of the 26th International Joint
  Conference on Artificial Intelligence. pp. 1725--1731 (2017)

\bibitem{Hu2015Thesis}
Hu, Z.: Initializing the EM Algorithm for Data Clustering and Sub-population
  Detection. Ph.D. thesis, The Ohio State University, Ohio, USA (Dec 2015)

\bibitem{Li2020prototypical}
Li, J., Zhou, P., et~al.: Prototypical contrastive learning of unsupervised
  representations. In: Proceedings of the 9th. International Conference on
  Learning Representation. ICLR '21 (2021)

\bibitem{Lin2022StEN}
Lin, S., Yu, Y., Ji, X., Zhou, T., He, H., Sang, Z., Jia, J., Cao, G., Hu, N.:
  Spatiotemporal-enhanced network for click-through rate prediction in
  location-based services. In: Proceedings of the 2022 ACM on Conference on
  Information and Knowledge Management. CIKM '22 (2022)

\bibitem{Sheng2021STAR}
Sheng, X.R., Zhao, L., Zhou, G., Ding, X., Dai, B., Luo, Q., Yang, S., Lv, J.,
  Zhang, C., Deng, H., Zhu, X.: One model to serve all: Star topology adaptive
  recommender for multi-domain ctr prediction. In: Proceedings of the 30th ACM
  International CIKM (2021)

\bibitem{sutton1999reinforcement}
Sutton, R.S., Barto, A.G.: Reinforcement learning: An introduction. Robotica
  \textbf{17}(2),  229--235 (1999)

\bibitem{Wang2019SRS}
Wang, S., Hu, L., Wang, Y., Longbing, C., Sheng, Q.Z., Orgun, M.: Sequential
  recommender systems: Challenges, progress and prospects. In: Proceedings of
  the Twenty-Eighth International Joint Conference on Artificial Intelligenc.
  IJCAI '19 (2019)

\bibitem{Zhou2018DIN}
Zhou, G., Song, C., Zhu, X., Fan, Y., Zhu, H., Ma, X., Yan, Y., Jin, J., Li,
  H., Gai, K.: Deep interest network for click-through rate prediction. In: The
  24th ACM SIGKDD Conference on Knowledge Discovery and Data Mining. KDD '18
  (2018)

\bibitem{zhou2018deep}
Zhou, G., Zhu, X., et~al.: Deep interest network for click-through rate
  prediction. In: Proceedings of the 24th ACM SIGKDD. pp. 1059--1068 (2018)

\end{thebibliography}

\end{document}